\documentclass[12pt]{iopart}

\usepackage{iopams}

\usepackage{epstopdf}
\usepackage{psfrag}
\usepackage{ps4pdf}

\usepackage{multirow}
\usepackage{paralist}
\usepackage{color}
\usepackage{graphicx}

\pdfoutput=1

\begin{document}

\newcommand{\D}{\mathrm{d}}
\newcommand{\pr}{{}^{\tiny{\mbox{ïð}}}}
\newcommand{\nuint}{\nu_{\mbox{\footnotesize{èíò}}}}
\newcommand{\Vreg}{V_{\mbox{\footnotesize{ðåã}}}}

\newcommand{\OB}[2]
{\overbrace{\hspace{#1mm}}^{ #2}}

\newcommand{\zd}
{
\scriptstyle{b-z(\tau)}\;
 \Bigg\{
}

\newcommand{\BC}{
\OB{22}{\frac{b}{c}}
}

\newcommand{\ZC}{
\OB{10}{\frac{b-z(\tau)}{c}}
}

\title{Comment on 'Simultaneous gravity and gradient
measurements from a recoil-compensated
absolute gravimeter'}
\author{V D Nagornyi}
\address{Axispoint, Inc., 350 Madison Avenue, New York, NY 10017, USA}
\ead{vn2@member.ams.org}
\begin{abstract}
The article (\emph{Niebauer et al.} 2011 \emph{Metrologia} \textbf{48} 154-163) reports on the important innovations enhancing the ability of absolute gravimeter to measure vertical gravity gradient along with the gravity acceleration. This comment suggests experiments to further assess the improvements and the results obtained with the modified instrument, considers some limitations of non-linear models in metrology and ways to overcome them, and discusses possible applications of the described instrument.
\end{abstract}
\newpage
\pagestyle{empty}
%
%
%
The paper \cite{niebauer2011a} reports on the important progress in gravimetric instrumentation. Following the efforts of \cite{hipkin1999,robertson2001}, the authors have made a new significant step in determining the vertical gravity gradient from the data obtained by absolute gravimeter. This comment offers some thoughts and suggestions on the interpretation of the reported results, and on the enhancements of the presented techniques.
\section*{Experimental results: reported and not reported}
Some findings presented in the paper  \cite{niebauer2011a} as the result of laborious  experimental efforts are actually known from theory. For example, the conclusion that the increased density of the scaled fringes does not affect the results, follows from the representation of ballistic gravimeter as low-pass filter \cite{svetlov1997}.
Another not totally unexpected conclusion is that the finite speed of light affects the measured gravity gradient. The trajectory disturbances caused by both components are well studied \cite{kuroda1991, timmen2003}, and their non-orthogonality is beyond doubt. At the same time, some important questions regarding the presented innovations, that can be answered only experimentally, were not addressed in the paper. They include
\begin{itemize}
\item How beneficial are the implemented modifications for the measurement of absolute gravity acceleration performed, for example, with the new absolute gravimeter
    \linebreak{FG5-X}
    \cite{FG5-X2011}? How does the gravity values measured before and after the modifications compare?
\item How does the absolute gravity value obtained with the non-linear model compare to that obtained with the traditional linear one?
\item How does the measured absolute value of gravity gradient compare to that measured in a traditional way by relative gravimeters?
\end{itemize}
\section*{Limitations of non-linear models}
The method of measuring the gravity acceleration and its vertical gradient, implemented in \cite{niebauer2011a}, uses least square fitting of the non-linear trajectory model\footnote{In the implemented model the non-linear parameters are $v_0$, $g_0$, and $\gamma$.}. Such models, however, possess several properties \cite{bard1973, dreiper1998} with unfavorable metrological interpretation.
\begin{itemize}
\item \emph{Bias.} The expectation of a parameter estimate by a non-linear model does not equal the expectation of the parameter itself. In other words, the parameter estimates are always biased in non-linear models. Sometimes the bias is so big that it can not be overlooked. The paper \cite{parker1995} reports the bias of 29~ $\mu$Gal, obtained by the non-linear fitting of the fringe signal. The same nonlinear model implemented in the laboratory testing of the complex heterodyne revealed a bias equivalent to 7~$\mu$Gal \cite{niebauer2006}.  Another paper \cite{bich2008} reports  the bias in the gradient of at least 3000~E, obtained by the model similar to that used in the paper \cite{niebauer2011a}. While decreasing with the data noise, the bias can not completely disappear. The questions posted in the previous section could help assessing the severity of this problem for the experiments conducted in \cite{niebauer2011a}.

\item \emph{Multiple minima.} For non-linear models, the surface of the squared residuals is not guaranteed to have a single minimum. The paper \cite{parker1995} gives an example of multi-minima surface. The results presented in the paper \cite{niebauer2011a} neither exclude the existence of this problem. Based on the slope of theoretical curve on the figure 10 of \cite{niebauer2011a}, we can estimate the uncertainty of the lead weights positioning, in case their setup is responsible for the repeatability error (fig.~\ref{fig_SETUP_ERROR}).
\begin{figure}[h]
\centering
\small
\includegraphics[height=60mm]{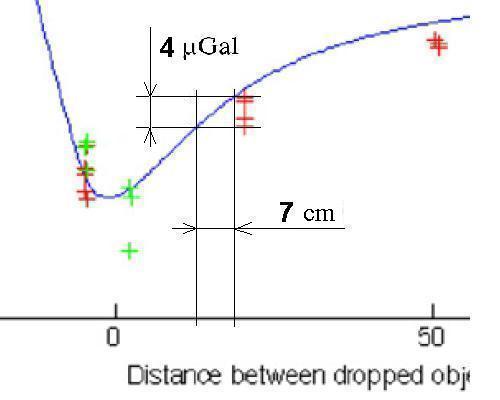}
  \caption[short title]
  {
  Fragment of the fig. 10 of \cite{niebauer2011a} discussing a possible lead weights position uncertainty as primary source of repeatability  error.
  }
\label{fig_SETUP_ERROR}
\end{figure}
It's difficult to admit that the platform with the weights could not be positioned better than with the 7 cm uncertainty. We therefore can assume that the repeatability error is partially explained by the existence of multiple minima in the squared residuals surface, and by the convergence of the iterative process to different minima in every measurement.
\end{itemize}
%
Evaluating of how the bias and the multiple minima affect the result is a difficult problem with no profound analytical or numerical  solution. Fortunately, the innovations presented in the paper \cite{niebauer2011a}, such as the extended trajectory and the reduced recoil, open up a wide range of possibilities to determine the gravity and its vertical gradient without resorting to non-linear models. The opportunities include fitting the linear model to different segments of the trajectory, manipulating the gradient input through the data re-sampling, adaptive linear filtering, and others.
\section*{Gravity gradient in measurement equation}
The ability of absolute gravimeters to autonomously determine the vertical gravity gradient can be very helpful for absolute gravity surveys. Though the methods to exclude the gradient from the measurement equation have long existed \cite{timmen2003}, some absolute gravimeters still use the gradient to obtain the gravity value \cite{niebauer1995, microglacoste2007}.  In this case, the gradient, according to the GUM (JCGM 100:2008) \cite{GUM2008}, acts as \emph{influence quantity} and its uncertainty must be included in the instrument's uncertainty budget. Despite the significant progress in determining the gradient presented in \cite{niebauer2011a}, the only information of its uncertainty is that ``the value for the gradient tends to the expected free-air gradient (3000E)'' \cite{niebauer2011a}.
Once the gradient uncertainty is available, the uncertainty of the absolute acceleration could be estimated.
Without the estimate, the figures 10 and 11 of \cite{niebauer2011a} experimentally confirm only the characteristics of the instrument as relative gravimeter and relative gradientometer.
\section*{Applications of the combined gravi-gradientometer}
The described in \cite{niebauer2011a} instrument has several advantages compared to the commercially available relative gravimeters, such as simultaneous determination of the gravity and its gradient, and no need for calibration. Possible applications of the instrument may include monitoring of volcanic activity \cite{dagostino2008}, experiments related to solar eclipses \cite{arnautov2009}, and validation of tidal models \cite{avsyuk2008}. The presented innovations are very promising for the ballistic measurements of the gravitational constant $G$ \cite{schwarz1998a}. Interested users will definitely appreciate the suggested in \cite{niebauer2011a} application of the instrument to monitor facilities like Fort Knox. To implement the idea, however, more steps seems desirable to proof the concept. The steps may include suggestions on reconfiguring the storage units, so that the monitored masses would be within the device sensitivity range; or even the idea of using the array of the devices could be considered to provide the full sensitivity coverage throughout the storage. These extra efforts are no doubt well justified by the prospects of being able to confidently detect, as shown in \cite{niebauer2011a}, the defection of as little as 280 kg of materials in just about 1 hour.
\section*{References}


%
\bibliographystyle{ieeetr}
\end{document}